\documentclass[prb,letterpaper,superscriptaddress,showpacs,floats,twocolumn]{revtex4}

\usepackage{graphicx}
\usepackage{amsfonts}
\usepackage{amsmath,amssymb}
\usepackage{hyperref}
\usepackage{hypernat}
\usepackage{color}

\definecolor{rltred}{rgb}{0.75,0,0}
\definecolor{rltgreen}{rgb}{0,0.5,0}
\definecolor{rltblue}{rgb}{0,0,0.75}
\hypersetup{colorlinks,%
hypertexnames=true,%
pdfauthor={},%
pdftitle={},%
pdfview=FitV,%
pdfstartview=FitV,%
pdfpagemode=UseNone,%
pdfpagelayout=TwoPageLeft,%
bookmarksopen=true,%
bookmarksnumbered=true,%
pdfhighlight=/I,%
linkcolor=rltred,%
citecolor=rltgreen,%
linktocpage=true}

\newcommand{\refeq}[1]{\autoref{#1}}

\newcommand{\appendixref}{\hyperref[sec:ana]{appendix~\ref*{sec:ana}}}

\newcommand{\unitv}[1]{\mathbf{\hat{#1}}}

\newcommand{\yzero}{y_{0}}
\newcommand{\lD}{\lambda_{\text{D}}}
\newcommand{\kF}{k_{\text{F}}}
\newcommand{\pF}{p_{\text{F}}}
\newcommand{\EF}{E_{\text{F}}}
\newcommand{\meff}{m_{\text{eff}}}
\newcommand{\heff}{h_{\text{eff}}}

\newcommand{\Areg}{A_{\text{reg}}}
\newcommand{\Ach} {A_{\text{ch}}}

\newcommand{\vreg}{v_{\text{reg}}}

\newcommand{\vch} {v_{\text{ch}}}
\newcommand{\hmin}{h_\text{min}}

\newcommand{\expval}[1]{\left\langle#1\right\rangle}

\newcommand{\cA}{\mathcal{A}}
\newcommand{\cT}{\mathcal{T}}
\newcommand{\Int}{\int\limits}
\newcommand{\ud}{\text{d}}

\newcommand{\aizm}{{z_m}}
\DeclareMathOperator{\Ai}{Ai}

\newcommand{\abs}[1]{\left|#1\right|}
\newcommand{\atxz}{\Big|_{x=0}}

\newcommand{\chinorm}[3][y]{\chi_{#2}^{#3}(#1)}
\newcommand{\chish}[1][y]{\chinorm[#1]{m}{\infty+}}
\newcommand{\chimwl}{{\chi_m^{\infty+}(w_L)}}
\newcommand{\chimnoarg}{{\chi_m^{\infty+}}}
\newcommand{\kxsh}[1][\infty+]{{k_{x,m}^{#1}}}
\newcommand{\Dkxm}{{\Delta k_{x,m}}}
\newcommand{\sigmy}[1][y]{{\sigma_m(#1)}}
\newcommand{\varm}{{\varepsilon_m}}

\newcommand{\aiintam}{{\mu_m}}
\newcommand{\aiintbm}{{\nu_m}}

\newcommand{\ie}{i.e., }
\newcommand{\Schro}{Schr\"o\-din\-ger }
\newcommand{\Poin}{Poincar\'e }
\newcommand{\PoinHus}{Poincar\'e-Husimi }
\newcommand{\eg}{e.g.\ }
\newcommand{\cf}{cf.\ }

\begin{document}

\title{Nano-wires with surface disorder:\\Giant localization lengths and
dynamical tunneling in the presence of directed chaos}

\author{J.~Feist}
\email{johannes.feist@tuwien.ac.at}
\affiliation{Institute for Theoretical Physics,
             Vienna University of Technology, 1040 Vienna, Austria}

\author{A.~B\"acker}
\author{R.~Ketzmerick}
\affiliation{Institut f\"ur Theoretische Physik, Technische Universit\"at
             Dresden, 01062 Dresden, Germany}

\author{J.~Burgd\"orfer}
\author{S.~Rotter}
\affiliation{Institute for Theoretical Physics,
             Vienna University of Technology, 1040 Vienna, Austria}

\date{\today}

\begin{abstract}
 We investigate electron quantum transport through nano-wires with
 one-sided surface roughness in the presence of a perpendicular magnetic field.
 Exponentially diverging localization
 lengths are found in the quantum-to-classical crossover regime,
 controlled by tunneling between regular and chaotic
 regions of the underlying mixed classical phase space.
 We show that each regular mode possesses a well-defined 
 mode-specific localization length.
 We present analytic estimates of these mode localization lengths
 which agree well with the numerical data. The coupling between
 regular and chaotic regions can be determined
 by varying the length of the wire leading to
 intricate structures in the transmission probabilities.
 We explain these structures quantitatively by dynamical
 tunneling in the presence of directed chaos.
\end{abstract}

\maketitle

\noindent

\section{Introduction}

\begin{figure}
  \centering
  \includegraphics[width=\linewidth]{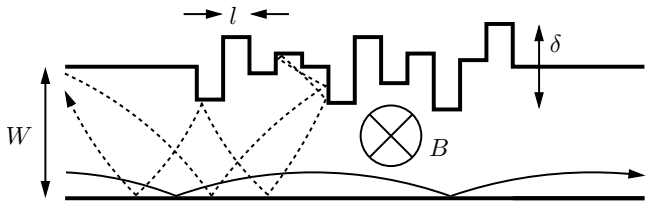}
  \caption{Wire with one-sided surface disorder in a magnetic field $B$,
  applied perpendicular to the scattering area. The solid line shows
  a regular skipping trajectory and the dashed line an irregular trajectory
  scattering at the disordered surface.}
  \label{fig:osdwire_sketch}
\end{figure}
\begin{figure}
  \centering
  \includegraphics[width=\linewidth]{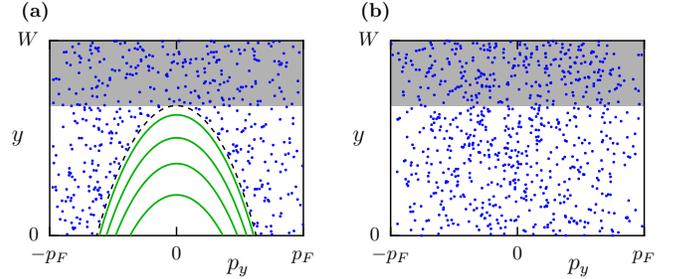}
  \caption{(Color online) \Poin sections for (a) right-moving ($p_x>0$) and (b) left-moving ($p_x<0$)
      classical electrons in the wire shown in \autoref{fig:osdwire_sketch}.
      In (a) a single regular island with invariant tori (green) can be identified,
      which is separated from the chaotic region (blue dots) by the outermost torus 
      (dashed). In (b), no such island appears.
      The gray-shaded part indicates the $y$-range affected by disorder, \cf\autoref{fig:section}.}
  \label{fig:osdwire_poincare}
\end{figure}

Deterministic dynamical systems with a mixed regular-chaotic phase space give rise to
many interesting features that are present in neither the regular nor
the chaotic limit alone.\cite{Gut1991,MarMey1974} These special features
include, in the purely classical limit, the emergence of \emph{directed
  chaos}, where the chaotic part of phase space exhibits a 
diffusive motion with a drift in a specific direction.\cite{Efe1997,FlaYevZol2000,%
SchOttKet2001SchDitKet2005,PruSch2006} 
Interestingly enough, such a biased diffusion may be realized in a
mixed phase space even without external driving, requiring only
broken time-reversal symmetry.\cite{SchPru2005} Quantized versions of mixed systems
also have interesting properties, since different regular and chaotic regions in phase space,
which are completely separated classically, become connected by
\emph{dynamical tunneling}.\cite{DavHel1981}
Dynamical tunneling can be viewed as a generalization of conventional tunneling
through potential barriers to multi-dimensional non-separable dynamical systems 
where the tunneling path along a specific ``reaction coordinate'' is, in general,
not well defined. Dynamical tunneling in phase space plays an important
role in spectral properties and transport.\cite{HanOttAnt1984,FisGuaReb2002,HufKetOtt2002,IomFisZas2002,%
BaeKetMon2005BaeKetMon2007,TomUll1994BohTomUll1993,PodNar2003,BaeKetLoe2008a,BaeKetLoe2008b}
Dynamical tunneling rates between regular regions are substantially enhanced by the presence
of chaotic motion, giving rise to the notion of chaos-assisted tunneling.\cite{TomUll1994BohTomUll1993}
Such dynamical tunneling processes were recently observed with cold atoms in periodically modulated
optical lattices.\cite{HenHafBro2001,SteOskRai2001}

In the present paper, we study transport in a long quantum wire
with one-sided surface disorder in the presence of a perpendicular homogeneous
magnetic field (see \autoref{fig:osdwire_sketch}). This system features
coexistent regions of regular motion (skipping trajectories) and irregular
motion induced by scattering at the disordered surface. In the
corresponding mixed phase space (\autoref{fig:osdwire_poincare}) we find
directed regular and irregular motion, which are quantum mechanically coupled
by dynamical tunneling. As a result, this scattering system gives rise 
to many interesting properties which are 
clearly reflected in the transmission through the wire---a quantity
which is readily accessible in a measurement. Surface disordered
wires and waveguides have recently received much attention, both
theoretically\cite{LeaFalLam1998,SanFreMar1999,GarSae2001,GarGovWoe2002,%
IzrMenLun2003,ChaSteNav2005} and experimentally.\cite{SonWanXia2005,KuhIzrKro2008,HonSohHwa2008}
This is mainly due to the fact that nano-wires, albeit being conceptionally
simple, are very rich in their physical properties. 
Nano-wires are now also being realized as graphene nano-ribbons,
for which surface disorder seems to play an
even more important role than for conventional (semi-conductor based)
wires.\cite{EvaZozXu2008,MucCasLew2009}

In a previous paper \cite{FeiBaeKet2006} we showed 
both numerically and analytically that by increasing
the number of open channels $N$ in the wire, or
equivalently, by increasing
the wavenumber $\kF$, the localization length $\xi$ induced by surface
disorder increases exponentially. 
The localization length was related to tunneling of the lowest
transverse wire mode ($m=1$) from the regular island to the chaotic region
in phase space (see \autoref{fig:osdwire_poincare}). The dramatic
increase of $\xi$ then follows
directly from the exponential suppression of the tunneling rates in
the semiclassical limit of large $\kF$ (or small de Broglie wavelength
$\lD$). In the present paper
we explore the behavior of higher transverse modes,
$m>1$. We give for each of these modes $m$ a remarkably
accurate analytical estimate for its specific localization length, $\xi_m$.
Furthermore, all modes in the regular island are effectively coupled to
one another by dynamical tunneling in the presence of irregular, yet directed,
motion: key is here the interplay between
directed regular motion (to the right in
\autoref{fig:osdwire_sketch}) and counter-moving irregular motion
directed to the left which gives rise to characteristic
structures in the mode-specific transmission probabilities $T_m$ of
the current-transporting
regular modes $m$. These intricate structures can be accounted for by a simple
scattering model incorporating opposing directed regular and irregular
motions and their coupling by tunneling.

This paper is organized as follows: In \autoref{sec:wire-osd} we briefly review the
characteristic classical and quantum features of our nano-wire, as 
induced by its mixed phase space. In \autoref{sec:loclen} we provide 
analytic estimates for the individual mode localization lengths
$\xi_m$. Technical details of
the underlying calculations are deferred to \appendixref.
In \autoref{sec:mode_coupling} we analyze the individual mode-to-mode
specific transmission probabilities $T_{m,m'}$, for which a conceptually
simple transport model based on the coupling by dynamical tunneling
between regular and irregular directed flow in opposite directions.
The paper is rounded off by a summary in \autoref{sec:summary}.

We use atomic units, but include the constants $\hbar
= \meff = e = 4\pi\varepsilon_0 = 1$ where instructive.

\section{Wire with surface disorder}\label{sec:wire-osd}

\subsection{Classical dynamics}\label{sec:class-prop}

We consider  a 2D wire with one-sided surface disorder to which two leads of 
width $W$ are attached (\autoref{fig:osdwire_sketch}). A homogeneous
magnetic field $B$ perpendicular to the wire is present throughout 
the system.
We choose the magnetic field to be directed in negative $z$-direction.
The Hamiltonian $H=\frac12 (\mathbf{p}+\mathbf{A})^2 + V(x,y)$ is then given in Landau gauge ($\mathbf{A}=By\,\unitv{x}$) by
\begin{align}\label{eq:land_ham}
	H &= \frac12 (p_x + B y)^2 + \frac{p_y^2}{2} + V(x,y) \,.
\end{align}
In the leads ($x<0$ and $x>Ll$)
\begin{subequations}
\begin{align}
  V(x,y) &= V_0 \left[ \Theta(-y) + \Theta(y-W)         \right] \,,\\
\intertext{and inside the quantum wire $0<x<Ll$}  
  V(x,y) &= V_0 \left[ \Theta(-y) + \Theta(y-W+\eta(x)) \right] \,.
\end{align}
\end{subequations}
$V_0$ is taken to be arbitrarily large ($V_0\to\infty$) to represent hard wall
boundary conditions. To emulate stochastic classical scattering at the
upper wire surface we choose $\eta(x)$ to be a random variable that is
piecewise constant for a fixed interval length $l$ (see
\autoref{fig:osdwire_sketch}). We
choose the value of $\eta(x)$ to be uniformly randomly distributed in
the interval 
\begin{equation}\label{eq:eta_range}
-\delta/2 \leq \eta(x) \leq \delta/2 \,.
\end{equation}
Thus, the wire is assembled from $L$ rectangular elements, referred to
in the following as modules, with equal width $l$, but random heights $h$,
uniformly distributed in the interval $[W-\delta/2,W+\delta/2]$.  
In the numerical computations we use $l=W/5$ and $\delta=(2/3) W$.
The Hamiltonian \autoref{eq:land_ham} is non-separable for a given
realization of disorder. It is therefore quite likely that generally
in such a system, mixed regular and chaotic motion
may ensue. In the following, we refer to the irregular motion in the wire as
``chaotic'', although we will only make use of the weaker
condition of stochasticity and ergodic coverage of the corresponding phase
space region due to the random disorder. The ensemble-averaged
value of the Lyapunov exponent does not enter the subsequent analysis.

The classical motion in the wire proceeds on circular arcs
characterized by the cyclotron radius
$r_c=\pF/B$ and guiding center coordinate $\yzero=-p_x/B$,
interrupted by elastic reflections
on the hard wall boundary. For $\yzero$ sufficiently small or negative such that
$\yzero+r_c \leq W-\delta/2$, the electron performs regular skipping motion, for
which $\yzero$ is conserved. These skipping trajectories generate a directed,
ballistic motion to the right (for the direction of the $B$-field and boundary 
conditions depicted in \autoref{fig:osdwire_sketch}). For trajectories hitting
the upper disordered boundary ($\yzero+r_c>W-\delta/2$), chaotic motion will
develop with an average drift to the left, as discussed below.

To visualize the classical dynamics for electrons with Fermi momentum $\pF$ 
we choose as \Poin section
a vertical cut at the entrance of the wire ($x\!=\!0$) 
with periodic boundary conditions in the
$x$-direction,\cite{SchPru2005} see \autoref{fig:osdwire_poincare}.
The resulting section ($y$, $p_y$) for $p_x> 0$ 
shows a large regular region with invariant
tori corresponding to the skipping motion along the lower straight boundary of
the wire. The section for $p_x<0$ shows only an irregular region, as all trajectories
with $p_x<0$ interact with the upper boundary.
The area $A$ in the \Poin section enclosed by a torus is given by 
\begin{equation} \label{eq:A}
  A=\pF r_c \left[\arccos (1-\nu) - (1-\nu) \sqrt{1-(1-\nu)^2}\right],
\end{equation}
where $\nu r_c$ is the $y$-position at the top of the corresponding
cyclotron orbit.  
The size $\Areg$ of the regular island is obtained for
$\nu_{\text{max}}=(W-\delta/2)/r_c$. Outside of the regular region the motion 
appears uniformly chaotic. Hierarchical structures of island chains are absent.

The system displays \emph{directed chaos},\cite{SchOttKet2001SchDitKet2005,SchPru2005,%
PruSch2006,BaeKetMon2005BaeKetMon2007} \ie the time averaged velocity of almost 
all classical trajectories in the chaotic region of phase space approaches 
a non-zero constant $\vch$ for long times. This chaotic drift motion
arises here as trajectories in the regular island 
have a non-zero average speed  $\vreg>0$
directed to the right, while the velocity average 
over the whole phase space is exactly zero,\cite{SchOttKet2001SchDitKet2005} $\Areg \vreg + \Ach \vch=0$.
Therefore the average velocity of the chaotic part must be directed to the left, $\vch<0$.

\subsection{Quantum description}
\begin{figure}
  \centering
  \includegraphics[width=\linewidth]{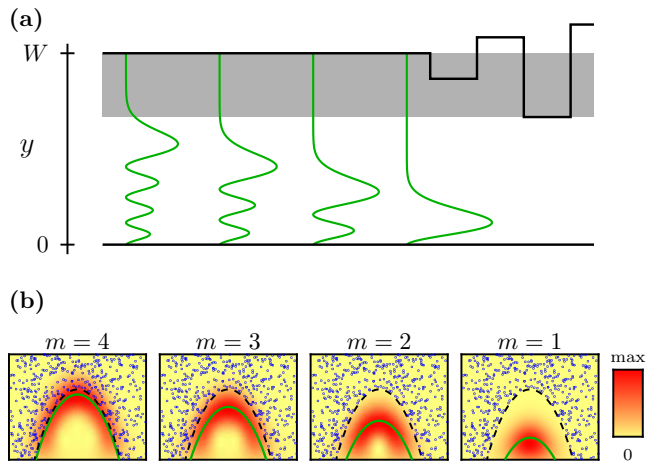}
  \caption{(Color online) 
      (a) Nano-wire with the regular transverse modes $\chi_m(y)$ 
      $m=4, 3, 2, 1$ (green) for $\kF W/\pi=14.6$.  The gray shaded part
      indicates the $y$-range affected by disorder.  
      (b) \PoinHus functions of these modes
      and their quantizing tori.}
  \label{fig:section}
\end{figure}

\begin{figure}
  \centering
  \includegraphics[width=\linewidth]{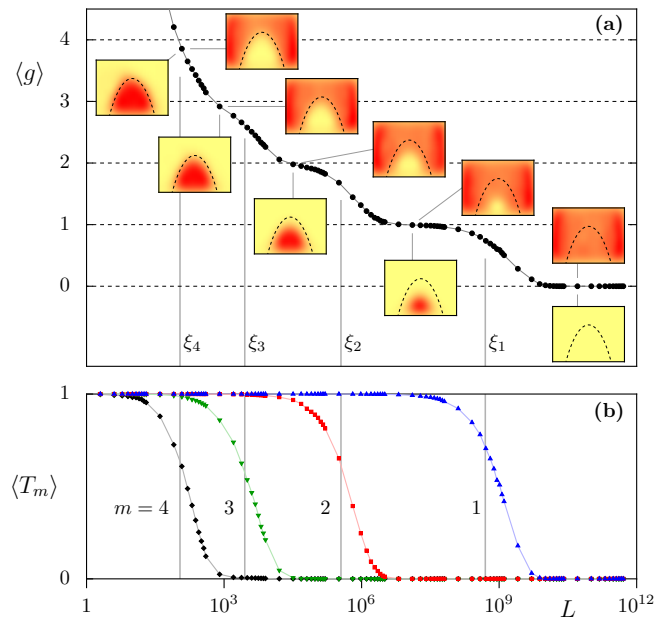}
  \caption{(Color online) (a) Averaged conductance $\expval{g}$
      vs~length $L$ of the wire for $r_c=3W$ and $\kF W/\pi=14.6$.  
      The step-wise decrease is accompanied by the disappearance
      of the regular modes (lower set of Husimi distributions) 
      and the flooding of the island 
      region by the chaotic states (upper set). The
      \PoinHus representations
      to the left (right) of the curve correspond to scattering
      from left to right (right to right) lead, respectively.
			(b) Transmission $\cT_m=\exp(\expval{\ln T_m})$ of the incoming mode $m$ vs~$L$.
			The light vertical lines correspond to the predicted localization
			lengths $\xi_m$, \autoref{eq:xis_asym}.}
  \label{fig:singlemodes}
\end{figure}

Quantum mechanically, the scattering through the wire is 
described in terms of the modes in the entrance and exit lead, respectively, 
see \autoref{fig:section}(a).
The lowest transverse modes of the incoming scattering wavefunctions
reside inside the regular island
(\autoref{fig:section}(b)). Only their exponential tunneling tail
in the harmonic-oscillator-like potential (see \autoref{eq:land_ham})
touches the upper disordered surface at $y>W-\delta/2$.
These regular modes can be semiclassically
quantized as\cite{KosLif1955,BeeHou1988}
\begin{equation} \label{eq:WKB}
  \frac{A}{h} = \frac{B \cA}{h/e} = (m-1/4) \quad \text{with $m=1,2,...$} \;\;,
\end{equation}
where $A$ is 
the area in the \Poin section enclosed by a given torus, see \refeq{eq:A},
and $\cA=r_c A /\pF$ is the area in position space enclosed by a segment of a
skipping orbit. 

The quantum states of the system can be represented
on the \Poin section by projecting
them onto coherent states  $\text{coh}_{(y,p_y)} (y')$,
which are localized in the point $(y,p_y)$,
\begin{equation}
  H_m(y,p_y)  =  \left| \int_0^{W} \chi_m(y')  \
                   \text{coh}_{(y,p_y)} (y') \; \ud y'  \right |^2 \;\;,
\end{equation}
to obtain a \PoinHus representation.
Unlike previous studies
(see [\onlinecite{BaeManHucKet2002,WeiRotBur2005BaeFueSch2004}] and references therein)
we use the states themselves and not their normal derivative
function as the section does not run along a border of
the billiard. The \PoinHus representation
for the lowest lead modes (\autoref{fig:section}(b)) clearly displays
the localization on the quantized tori.

As this system is a non-separable two-degree of freedom system, the 
one-dimensional states in the transverse direction $y$ do not remain
stationary in the diamagnetic harmonic-oscillator potential of the Hamiltonian in \autoref{eq:land_ham} (truncated at
$y=0,\eta$). Due to their extension to the rough upper wall, the
transverse states behave as an equivalent time-dependent one-degree 
of freedom system in the $y$-coordinate with a time-dependent potential
due to stochastic variations of the upper wall. Accordingly, each mode has
a characteristic decay time $\gamma_m$ which is controlled by the overlap of 
the tunneling tail with the rough wall.
Modes with small $m$ live near the center of the island and
therefore only couple weakly having small tunneling rates, while
states near the border of the island have stronger overlap with the disordered 
wall and consequently have large tunneling rates.

Numerically, the
particular realization of disorder allows for an efficient
computation of quantum transport for remarkably long wires  by 
employing the modular recursive Green's function method.\cite{RotTanWir2000,RotWeiRoh2003}
We first calculate the Green's functions for $M\!=\!20$ rectangular modules with 
different heights. A random sequence of these modules is connected by means of a matrix
Dyson equation.  Extremely long wires can be reached by implementing an
``exponentiation'' algorithm\cite{SkjHauSch1994}: Instead of connecting the
modules individually, we iteratively construct different generations of
``supermodules'', each consisting of a randomly permuted sequence of $M$
modules of the previous generation. Repeating this process leads to the
construction of wires with lengths that are exponentially increasing with
the number of generations.
With this approach we can study wires with up to $\sim
10^{12}$ modules, beyond which 
limit numerical unitarity deficiencies set
in.  For wires with up to $10^5$ modules we can compare this supermodule
technique containing pseudo-random sequences with truly random sequences of
modules.  For configuration-averaged transport quantities the results are
indistinguishable from each other even in the semiclassical limit of small
de Broglie wavelength, $\lD<l$, to be explored in the following. Accordingly,
we can simulate quantum wires with a length of the order of $\sim
10^{12}\lD$, which illustrates the remarkable degree of numerical
stability of the MRGM.

The transmission $t_{mn}$ and reflection amplitudes $r_{mn}$ for an electron
injected from the left are evaluated by projecting the Green's function at the
Fermi energy $\EF$ onto all lead modes $m,n\in \{1,\ldots,N \}$ in the
entrance and exit lead, respectively.  Here $N=\lfloor \kF W/\pi \rfloor$ is
the number of open lead modes and $\kF$ the Fermi wave number.  
From the transmission amplitudes one obtains the dimensionless
conductance $g={\rm Tr} (t^{\dagger} t)$.
The ensemble-averaged conductance $\expval{g}$ for 20 different disorder
realizations and three neighboring values of the wavenumber $\kF$ (\autoref{fig:singlemodes})
initially decreases sharply (not shown) since the contribution of modes with high $m$
($m>5$ in the present case), residing primarily in the chaotic sea and
transporting, on average, to the left, rapidly vanishes with increasing $L$.

For larger lengths $L$ the conductance shows
a number of steps with increasing length, which can be understood
as follows:
At first, the entire regular region
contributes to the conductance with 4 modes.
With increasing $L$, fewer and fewer regular modes contribute
to the conductance, because they decay due to tunneling.
This happens first for the outermost regular mode,
which has the largest tunneling rate $\gamma_m$.
Finally, at very large lengths of the wire, even
the innermost regular mode no longer contributes to the transport,
so that the conductance decays to zero.
In this limit and on this length scale, one-sided disorder of the
quantum wire leads to localization. 

The coupling to the 
chaotic sea by dynamical tunneling can also be viewed as 
``flooding'' of the regular island by chaotic states.\cite{BaeKetMon2005BaeKetMon2007}
This notion is illustrated by the \PoinHus distributions for states 
injected from and ejected to the right (see \autoref{fig:singlemodes}, discussed below).
The decay of the regular states into the chaotic sea or the reversed process,
penetration of the chaotic states into the regular island implies that the 
regular states are quasi-discrete states embedded in a (quasi-)continuum.
For closed wires (with periodic boundary conditions at $x=0$ and $x=L$) the discrete energy
levels which form the
quasi-continuum must feature a sufficiently
high density of states, or small level-spacing $\Delta$\,,
\begin{equation}\label{eq:newcondition}
\gamma_m > \Delta \,,
\end{equation}
for effectively quenching the quasi-discrete regular state
on the $m$-th torus and giving rise to chaotic eigenstates
flooding this torus. When increasing $L$ and
thus decreasing the mean level spacing, $\Delta \sim 1/L$,
this happens at characteristic lengths of the closed system
where $\gamma_m=\Delta$.

However, \autoref{eq:newcondition} is not applicable to open systems,
The step-wise behavior observed in \autoref{fig:singlemodes} is 
therefore determined by the different 
mode-specific tunneling rates $\gamma_m$ themselves, rather than by a
possible violation of \autoref{eq:newcondition}.

For the present scattering system the successive 
flooding of the regular island and the resulting
stepwise behavior of the conductance can be visualized
by considering averaged \PoinHus distributions gained
from random  superpositions 
of all modes entering from the left and scattering to the right
(insets below the curve of $\expval{g}$ in \autoref{fig:singlemodes}(a)).
One clearly sees how the contribution to the transport
from the regular region disappears with increasing length of the wire.
Also shown are the corresponding pictures obtained for 
scattering from right to right. For small $L$
these \PoinHus functions are first outside
the regular island which for increasing 
$L$ is flooded by the chaotic states. 
For the largest $L$ we have complete flooding,
\ie no regular modes are left and the chaotic modes
fully extend into the regular island.
Thus the disappearance of regular states
and the flooding of the island by chaotic states
is nicely seen in the complementarity of the Husimi pictures.

\section{Localization lengths}\label{sec:loclen}

We obtain quantitative results for the
localization length $\xi$ (in units of $l$) in a wire consisting of $L$ modules 
by analyzing the conductance $g$
in the regime $g \ll 1$, extracting $\xi$ from $\expval{\ln g}\sim -L/\xi$.  

In addition, we can determine mode-specific localization lengths for
the regular modes $m$: To this end, we consider 
 the individual transmission probabilities
$T_m=\sum_{n} |t_{nm}|^2$
for the incoming mode $m$ as a function of $L$, see
\autoref{fig:singlemodes}(b), where an average $\cT_m =\exp(\expval{\ln T_m})$ 
over 20 disorder realizations and three neighboring $k_F$-values is shown.
For each of the regular modes we observe an exponential decay subsequent to the diffusive regime.
Consequently, it is possible to define the
\emph{mode localization lengths} $\xi_m$. 
They can be obtained numerically by fitting 
to $\expval{\ln T_{mm}} \propto -L/\xi_m$.
The largest mode localization length is $\xi_1$, as
the corresponding regular mode $m=1$ couples most
weakly to the chaotic region.
Thus $T_1$ determines the conductance for long wires $L>\xi_1$
such that $\xi=\xi_1$.
The regular modes with $m>1$ contribute to the conductance 
for $L<\xi_m$, leading to the step-wise behavior 
observed in  \autoref{fig:singlemodes}.

We want to explore these localization lengths 
in the quantum-to-classical crossover.
To this end we introduce the semiclassical parameter $\heff$,
the ratio of Planck's constant to the area of the \Poin section,
\begin{equation}
	\heff = \frac{h}{2\pF W} = \left( \frac{\kF W}{\pi} \right)^{-1} ,
\end{equation}
which coincides with the inverse number of modes. 
We study the semiclassical limit, 
$\heff \to 0$ or equivalently $\kF\to\infty$,
for two different cases:

(i) The cyclotron radius $r_c$ is kept fixed by adjusting the magnetic field 
$B=\hbar \kF/ r_c$ for increasing $\kF$. 
This leaves the classical dynamics invariant. In particular,
the fraction of the regular phase space volume stays constant,
while the absolute size $\Areg$ of the regular island 
scales as $\Areg \propto \kF \propto \heff^{-1}$. 
At the same time, $\lD$ decreases and approaches the semiclassical limit.
In the numerical computations we use the value $r_c=3W$.
The case $m=1$ was originally studied in our previous paper.\cite{FeiBaeKet2006}

(ii) The magnetic field $B$ is kept fixed. Since the cyclotron radius $r_c= \hbar \kF/ B$
increases for increasing $\kF$, the electrons follow increasingly straight
paths, thus reaching the disordered surface more easily.
Therefore the fraction of the regular phase space volume decreases.
As can be deduced from \refeq{eq:A} in the limit $\nu \to 0$,
the absolute area $\Areg$ still increases,
$\Areg \propto \sqrt{\kF} \propto \heff^{-1/2}$, 
but much slower than in the previous case. 
In the numerical computations we use the value $B=10.05\,$a.u.
corresponding to $r_c=3W$ for $\heff^{-1}=9.6$.

\begin{figure}
  \centering
  \includegraphics[width=\linewidth]{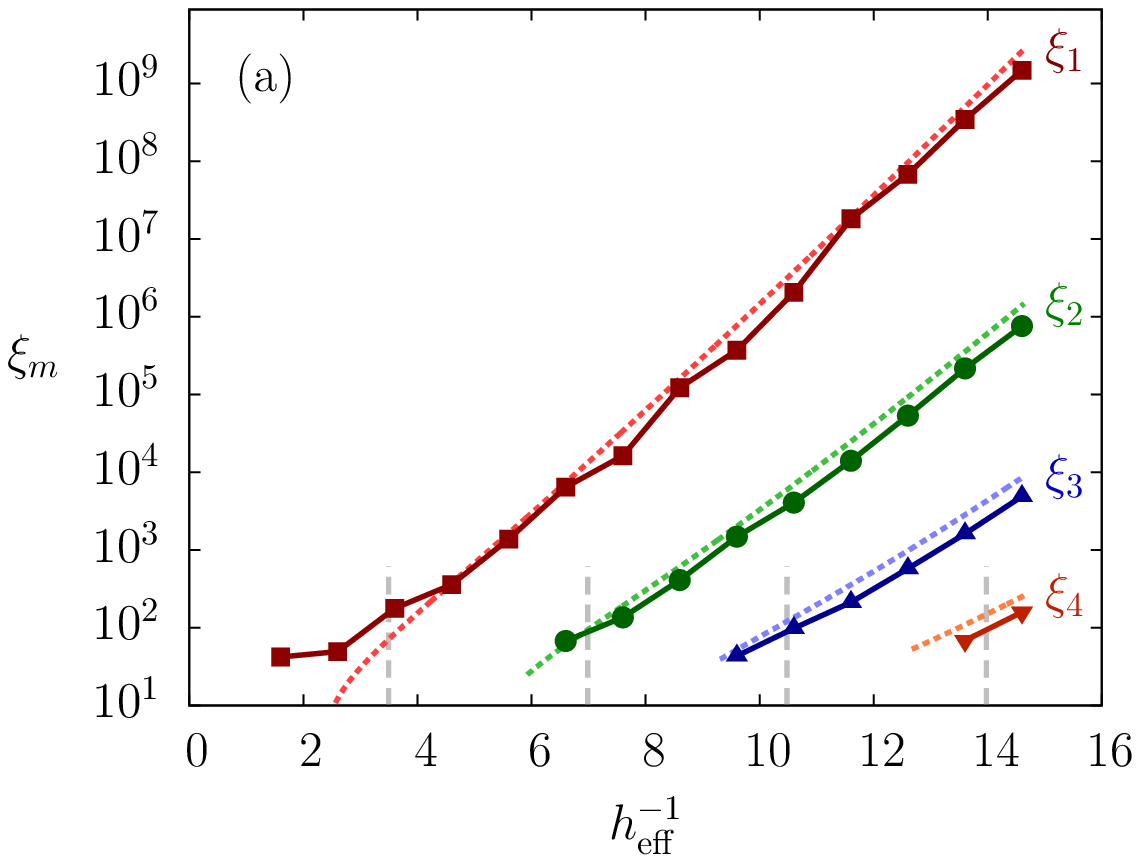}
  \includegraphics[width=\linewidth]{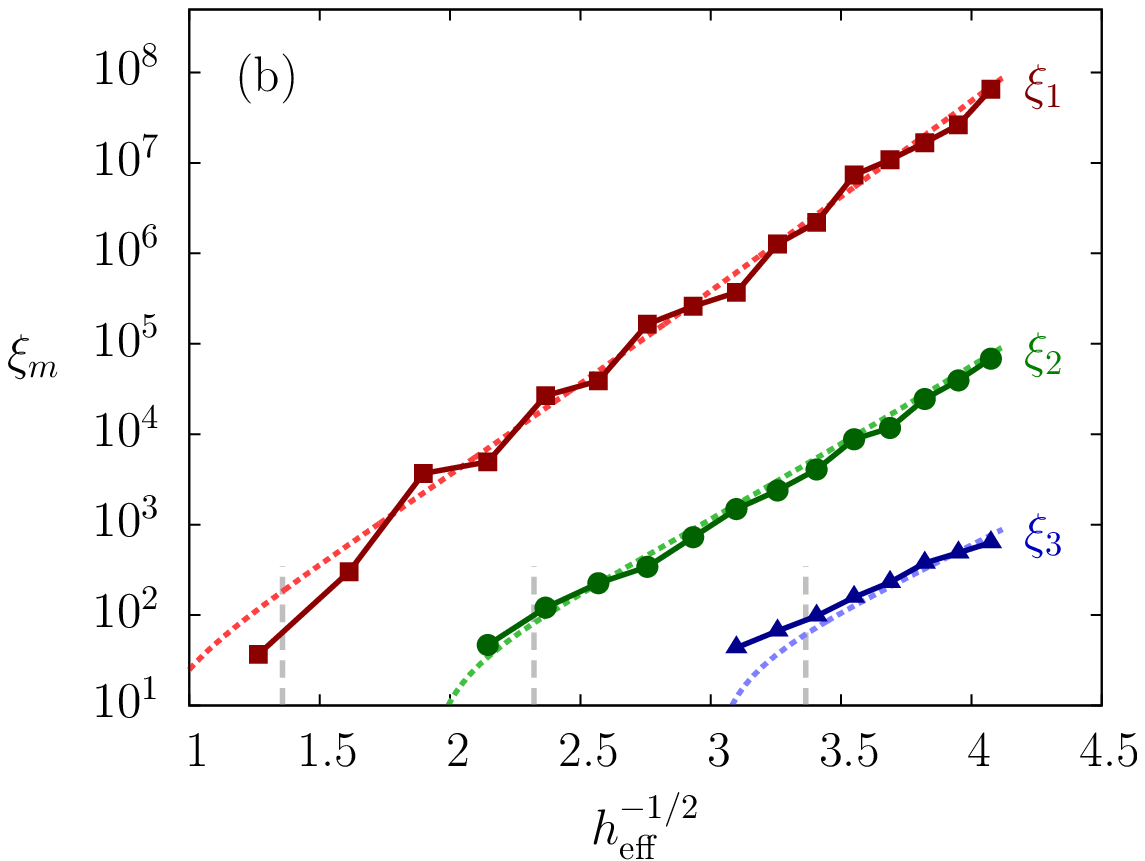}
	\caption{(Color online) (a) Mode localization lengths $\xi_m$ for constant cyclotron radius. 
	(b) Mode localization lengths $\xi_m$ for constant magnetic field $B$.
	The symbols connected by full lines display the numerical results from
	the full calculation.
	The dashed lines show the analytical predictions from (a) \autoref{eq:xis_asym} 
	and (b) \autoref{eq:final_constb}, while the vertical gray lines indicate the positions where
	the size of the regular island is large enough to accommodate $m$ modes (\ie where $\Areg=mh$).}
  \label{fig:xis_osd_b_n}
\end{figure}

\subsection{Qualitative description}

We start with a qualitative description that gives insight in the overall
dependence of the localization lengths for the two cases above.
It is based on the observation that tunneling from the central mode,
$m=1$, of a regular island to the chaotic sea
can be approximately described as being exponentially small in the ratio
of island size to Planck's constant,\cite{HanOttAnt1984, PodNar2003, BaeKetLoe2008a}
\begin{equation} \label{eq:gammaestimate}
  \gamma_1 \sim \exp\left(- C \, \frac{\Areg}{h} \right),
\end{equation}
with a system-dependent constant $C$.

Invoking again the equivalent description of a time-dependent one-degree of freedom system,
its temporal decay $\exp(-\gamma_1 t)$ 
leads to an exponential decay, $\exp(-\gamma_1 x/v_1)$, as a function of propagation length $x$.
This gives a localization length $\xi_1 \sim \gamma_1^{-1}$ [\onlinecite{HufKetOtt2002,IomFisZas2002}],
which in the limit of $\heff \to 0$ shows the following behavior,
\begin{align}
  \text{(i) $r_c$ fixed: } &\quad \xi_1 \sim \exp\Bigl(c_0\,\heff^{-1}   \Bigr)\,, \label{eq:xiestimate-i} \\
  \text{(ii)  $B$ fixed: } &\quad \xi_1 \sim \exp\Bigl(c_0\,\heff^{-1/2} \Bigr)\,, \label{eq:xiestimate-ii}
\end{align}
where the constant $c_0$ is different for each case and determined below.
Checking with our numerical results (full lines in \autoref{fig:xis_osd_b_n}) we find that,
on a qualitative level, 
these estimates correctly predict an exponentially increasing localization length $\xi_1$.
Also displayed are the localization lengths for higher modes $\xi_m$,
which are smaller for larger $m$. 
The onset of their exponential increase with $\heff \to 0$ can be linked 
to the critical size of the island such that its area is large 
enough to accommodate $m$ modes, $\Areg \approx mh$. 
From this relation one can determine the corresponding values of $\heff$.
In case (i) one finds for the numerically used parameters $\heff^{-1} \approx 3.5 m$,
in case (ii) one finds $\heff^{-1}\approx 0.32+0.59(m^2+m\sqrt{1.08+m^2})$.
These values are shown in \autoref{fig:xis_osd_b_n} 
by the dashed vertical lines.

The above analysis links the exponential increase of the localization
length to the existence of the island of regular motion.
An immediate consequence is that 
if such an island does not exist, as e.g.\ for two-sided disorder
or for $B=0$, an exponentially increasing localization length
is absent.\cite{FeiBaeKet2006} For wires with two-sided disorder
large localization lengths do, however, reappear in the quantum Hall regime 
where the cyclotron radius is much smaller than the wire width,
$r_c\ll W$, and thus much smaller than considered throughout this paper. 

\subsection{Quantitative description}

We now go beyond the above qualitative reasoning and derive 
analytical estimates for the localization
lengths $\xi_m$ of the island modes. For the present realization of disorder,
transitions between modes only occur at the boundaries between adjacent
modules of differing height. At each boundary the wave
functions to the left and to the right of the discrete jump have to be matched. 
The corresponding matching conditions can
be drastically simplified by considering that the lowest transverse 
modes in each module will differ only slightly from the corresponding
modes in the neighboring module of different height. This is because the
effective quadratic potential induced by the magnetic field, \autoref{eq:land_ham},
plays the role of 
a tunneling barrier through which only the evanescent part of the transverse
modes may reach the upper waveguide boundary where random fluctuations of module
heights occur. The evanescent part of the wavefunction can be represented by
a WKB-approximation for the corresponding tunneling integral. 
Within this semiclassical description, we obtain an approximate analytical
expression for the transmission
coefficients and, consequently, for the localization lengths. 
The details of this derivation are given in \appendixref.  We find for case (i) (fixed $r_c$)
\begin{equation}\label{eq:xis_asym}
  \xi_m \approx \left(a_m \, \heff^{-2/3} - b_m \right) 
        \exp \left[c_0 \, \heff^{-1} (1-d_m \, \heff^{2/3})^{3/2} \right] ,
\end{equation}
and for case (ii) (fixed $B$)
\begin{equation}\label{eq:final_constb}
  \xi_m \approx \left(a_m \, \heff^{-1/3} - b_m \right) 
        \exp \left[c_0 \, \heff^{-1/2} (1-d_m \, \heff^{2/3})^{3/2} \right] ,
\end{equation}
where in each case the constants $a_m, b_m,$ and $d_m$ depend on $m$,
while $c_0$ is independent of the mode number (see \appendixref{} for details).
The above predictions for the mode localization lengths $\xi_m$ are in
excellent agreement with the numerically obtained values (\autoref{fig:xis_osd_b_n}).
Moreover, the leading order dependence on $\heff$ is identical to the
qualitative predictions in \refeq{eq:xiestimate-i} and \refeq{eq:xiestimate-ii},
respectively.

\section{Mode-to-mode transition probabilities}\label{sec:mode_coupling}
Additional insights into the interplay between directed regular motion to
the right, directed chaotic motion to the left, and chaos-assisted tunneling 
can be gained from the mode-to-mode transition
probabilities $T_{m'm} = |t_{m'm}|^{2}$. These display intricate structures
as a function of $L$ due to the tunneling transitions between the
counter-propagation currents the details of which we investigate below. 

\begin{figure}
  \centering
     \includegraphics[width=\linewidth]{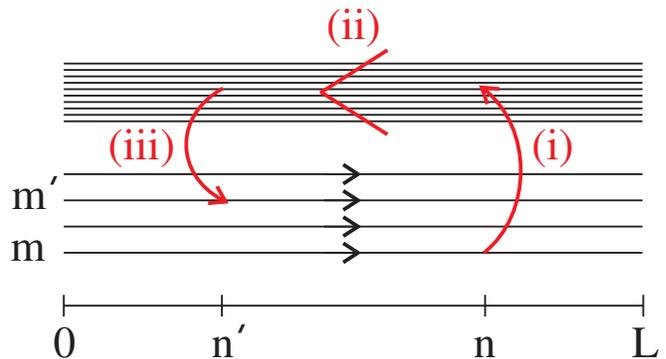}
  \caption{(Color online) Three-step process for transmission from regular mode $m$ 
  at module boundary $n$ to the left-transporting chaotic modes and finally to the
  regular mode $m'$ at the module boundary $n'$.}
  \label{fig:threestepsketch}
\end{figure}

The diagonal transmission of a given mode into itself is given to lowest order by
$T_{mm} = \exp(-L/\xi_m)$. 
The dominant contribution to the off-diagonal transmission probabilities 
$T_{m'm}$ with $m' \neq m$,
can be constructed based on the following three-step process (\autoref{fig:threestepsketch}):
(i) Tunneling from the right-moving regular mode $m$ to the chaotic sea,
(ii) propagation in the chaotic sea, which has an average drift to the left, 
and (iii) tunneling into the right-moving regular mode $m'$.
This three-step process incorporates all the basic elements of chaos-assisted tunneling,\cite{TomUll1994BohTomUll1993}
however here in the presence of directed regular and chaotic transport. 
In addition, the signature of chaos-assisted tunneling is here analyzed not
in terms of spectral properties, but in the transmission properties of an open systems.

\subsection{Qualitative analysis}

\begin{figure}
  \centering
  \includegraphics[width=\linewidth]{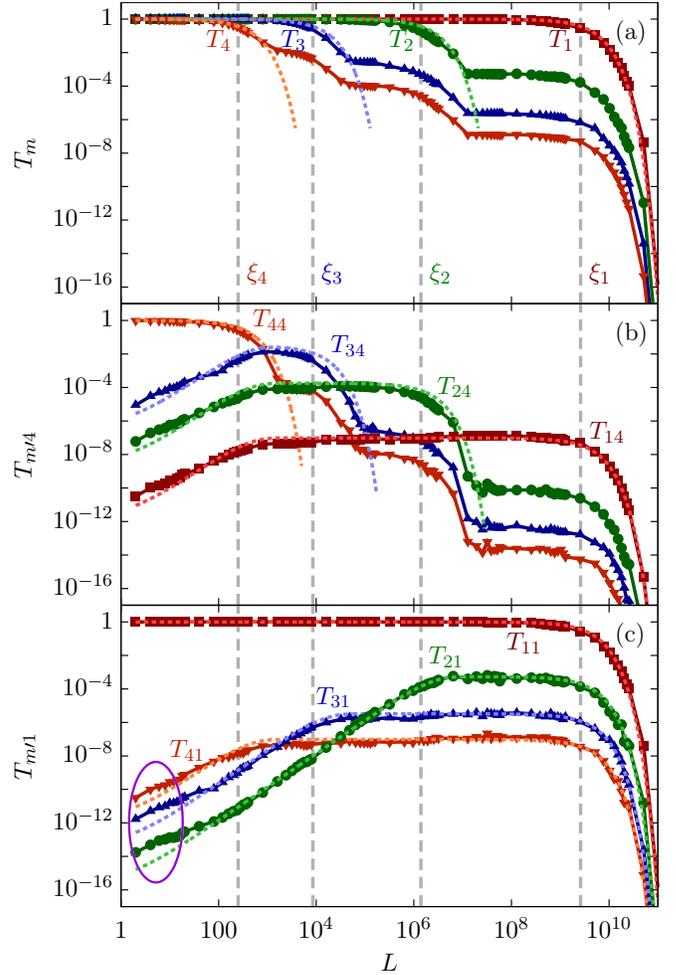}
  \caption{(Color online)
    (a) Transmission probabilities $T_m$ of the incoming mode $m(=1,2,3,4)$ vs.~$L$
    showing the same data as in \autoref{fig:singlemodes}(b), but on a 
    logarithmic scale.
    (b) Transmission probabilities $T_{m'4}$.
    (c) Transmission probabilities $T_{m'1}$.
    The dashed lines show the predictions from \autoref{eq:tm1est}, using 
    the analytical results for the localization lengths $\xi_m$.
    Second-order processes (\ie coupling from mode $m$ to $m'$ and then from $m'$ to $n$) are also taken into
    account.    
    For the numerical results, the arithmetic means $\expval{T_{m}}$ and $\expval{T_{m'm}}$ are taken.
    In regions where the distribution of transmission probabilities is log-normal, the geometric mean
    (\ie arithmetic mean of the logarithms) could be more appropriate. For simplicity we use
    the arithmetic mean everywhere. The agreement between analytical and numerical results is slightly
    better with arithmetic averaging.
    The purple ellipse indicates the region where the direct transition between island modes dominates (see text).
  }
  \label{fig:transmissions}
\end{figure}

The study of mode-to-mode transmission probabilities as a function
of the length $L$ gives insight analogous to a time-dependent
observation of a wave packet.
Starting point are the transmission probabilities $T_m = \sum_{m'} T_{m'm}$ 
for each regular mode $m$, now plotted on a logarithmic scale (\autoref{fig:transmissions}(a)), 
which were previously shown on a linear scale (\autoref{fig:singlemodes}(b)).
We find that the transmission probabilities $T_m$ which give rise to the stepwise decay 
of the total transmission $T$ also display a stepwise decay on their own:
For short wires, the $T_m$ decay exponentially with $\exp(-L/\xi_m)$. Beyond
$L \approx \xi_m$ a plateau is reached, followed by a slower decay $\exp(-L/\xi_{m-1})$.
This sequence continues until $T_m$ finally decays with 
the largest localization length $\xi_1$.

In order to explain these structures, we first consider the
case of $T_4 = \sum_{m'}  T_{m'4}$ (the lowest curve in \autoref{fig:transmissions}(a)).
Its individual contributions $T_{m'4}$ 
are shown in \autoref{fig:transmissions}(b).
With increasing $L$ the dominant contribution switches from
$T_{44}$ to $T_{34}$ to $T_{24}$ to $T_{14}$ 
at the localization lengths $\xi_4$, $\xi_3$, and $\xi_2$, respectively.
The appearance of the contributions $T_{34},T_{24},T_{14}$
can be well accounted for by the three-step model mentioned above (details given below).
The subdominant plateaus occurring in \autoref{fig:transmissions}(a) and (b)
are due to higher order effects and will be discussed at the end of the section.

\begin{figure}
  \centering
  \includegraphics[width=\linewidth]{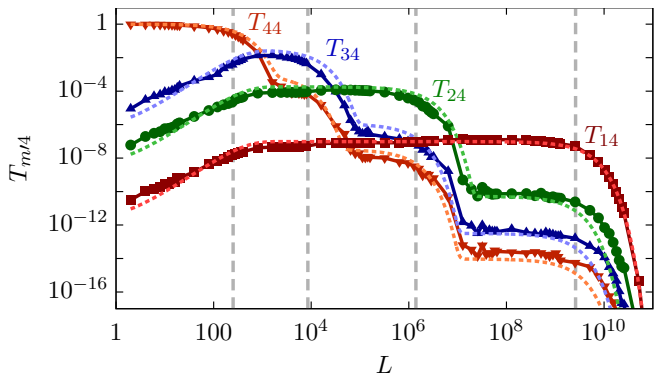}
  \caption{(Color online)
    Transmission probabilities $T_{m'4}$ as in \autoref{fig:transmissions}(b), but with
    the dashed lines showing the predictions from \autoref{eq:tm1est} with second-order 
    processes (\ie coupling from mode $m$ to $m'$ and then from $m'$ to $n$) also taken into
    account.}
  \label{fig:transmissions_2ndorder}
\end{figure}

The transmission of the innermost mode, $T_1= \sum_{m'}  T_{m'1}$ (shown 
at the top in \autoref{fig:transmissions}(a)) and of its individual contributions $T_{m'1}$ 
(shown in \autoref{fig:transmissions}(c)) displays a qualitatively different behavior. Still, it can also
be accounted for by the same three-step model: The transmission of mode $m=1$
is always dominated by the diagonal $T_{11}$, as it has the largest localization length.
For short wires the transmission probability $T_{21}$ is the smallest of the off-diagonal contributions,
as in step (iii) the mode $m'=2$ couples most weakly to the chaotic sea.
However, coupling becomes efficient with increasing wire length, leading to a higher saturation level
than $T_{31}$ and $T_{41}$. The saturation can be understood
as an equilibrium of the rate of populating mode $m'$ in step (iii)
via the chaotic sea and the rate of depopulation due to
the finite localization length $\xi_{m'}$ (step (i)).
The value of the saturation plateau will be determined from a quantitative analysis.

\subsection{Quantitative analysis}

We now demonstrate that dynamical tunneling 
in the presence of directed chaotic motion
allows for a quantitative description of the transmission probabilities. 
While the opposite direction of chaotic transport in step (ii) was not essential for
the above qualitative understanding of the transmission probabilities,
it is quantitatively of great importance.
The three steps of chaos-assisted tunneling (\autoref{fig:threestepsketch}), 
can be combined to give the following
estimate for $T_{m'm}$ with $m' \neq m$,
\begin{equation}\label{eq:tm1est}
   T_{m'm}(L) \approx \sum_{n=0}^{L} \sum_{n' \leq n}
             T_{mm}(n) \frac{1}{\xi_m} \frac{1}{\xi_{m'}} T_{m'm'}(L\!-\!n') \,,
\end{equation}
where the terms are related to the three-step process in the following way:

(i) The first tunneling process from mode $m$ to the chaotic sea can take place
at any module boundary $n$ along the wire consisting of $L$ modules.
The probability to remain in the initial mode $m$, before tunneling, is given by
$T_{mm}(n)$. The probability of tunneling into the chaotic sea
is proportional to the inverse localization length $\xi_m^{-1}$.

(ii) The chaotic sea features an average drift to the left. This is
incorporated in \autoref{eq:tm1est} by the restriction $n' \leq n$, 
where $n'$ is the location of the second tunneling process.
We do not distinguish explicitly between modes of the chaotic sea,
as they are strongly coupled and mirror the ergodicity of the underlying
classical dynamics.

(iii) The second tunneling process at $n'$ from the chaotic sea 
to the regular mode $m'$ has a probability proportional to 
the inverse of the localization length $\xi_{m'}$.
The last factor in \autoref{eq:tm1est}, $T_{m'm'}(L-n')$, 
describes the transmission probability within mode $m'$ from the module boundary
$n'$ to the exit lead at length $L$.

We now approximate the diagonal transmission probabilities 
$T_{mm}$ by their leading-order behavior, $T_{mm}(L)=\exp(-L/\xi_m)$. 
For $L,\xi_{m},\xi_{m'} \gg 1$ this results in
\begin{equation}\label{eq:tm1eval}
   T_{m'm}(L) \approx \frac{\xi_{m'} e^{-\frac{L}{\xi_{m}}} - \xi_{m} e^{-\frac{L}{\xi_{m'}}}}{\xi_m - \xi_{m'}}
        +  e^{-L\left( \frac{1}{\xi_{m}} + \frac{1}{\xi_{m'}} \right)} \,.
\end{equation}
\autoref{eq:tm1eval} is symmetric with respect to mode interchange, $T_{m'm}(L) \approx T_{mm'}(L)$.
Three limiting cases can be deduced, as
\begin{equation}
  \label{eq:tm1cases}
  T_{m'm}(L) \approx \left\{
	\begin{array}{lrl}
    \frac{L^2}{2\xi_m\xi_{m'}}   &   L &\ll \xi_m,\xi_{m'} \\ &&\\
    \frac{\xi_{m'}}{\xi_{m}} \exp\left(-L/\xi_{m}\right) \;\; & \xi_{m'}&\ll L, \xi_{m} \\ &&\\
    \frac{\xi_{m}}{\xi_{m'}} \exp\left(-L/\xi_{m'}\right) & \xi_{m} &\ll L, \xi_{m'}
    \,.
  \end{array}
  \right.
\end{equation}

These quantitative predictions involving chaos-assisted tunneling in the presence of 
directed chaotic motion are shown in \autoref{fig:transmissions},
demonstrating excellent agreement.
Here we have used the analytic predictions for the localization lengths,
given in the previous section. Thus, this agreement is achieved without any adjustable parameter.
We emphasize that without explicit use of directed chaotic motion,
e.g.\ without a restriction on $n'$ in \autoref{eq:tm1est}, one would get
drastically different predictions not compatible with the numerical results.
This confirms the notion of chaos-assisted dynamical tunneling in the presence of 
directed chaos.

Corrections to the transmission probabilities (\autoref{eq:tm1eval}) can be analyzed as well.
The direct tunneling between regular modes (\ie without a detour to the chaotic sea)
adds a contribution of the form
\begin{equation}\label{eq:tmm_direct}
   T_{m'm}^\text{direct}(L) \approx \sum_{n=0}^{L} 
   T_{mm}(n) \frac{\alpha_{m'm}}{\xi_m\xi_{m'}} \, T_{m'm'}(L-n) \,,
\end{equation}
where the factor $\alpha_{m'm}/(\xi_m\xi_{m'})$ denotes the direct 
transmission probability from mode $m$ to mode $m'$ at 
a single module boundary. Numerical results give a 
factor of the order of $\alpha_{m'm} \approx 10$, with small dependence on other parameters.
Only for very short wires, $L<2\alpha_{m'm}\ll\xi_m,\xi_{m'}$, \autoref{eq:tmm_direct} gives the
dominant contribution to the transmission $T_{m'm} \approx T_{m'm}^\text{direct}(L) \approx L\alpha_{m'm}/(\xi_m\xi_{m'})$,
linear in $L$, which can indeed be observed in \autoref{fig:transmissions}(c), indicated by the purple ellipse.

Higher order tunneling processes explain the subdominant plateaus in
\autoref{fig:transmissions}(b). For example $T_{34}(L)$ displays, after the
first plateau and the
exponential decay $\exp(-L/\xi_3)$, further plateaus related to
chaos-assisted tunneling from mode $m=4$ to mode $m''$ and then another
chaos-assisted tunneling from mode $m''$ to mode $m'=3$. In general this gives
for $T_{m'm}$ with $m \geq  m'> m''$
plateau values $(\xi_{m} / \xi_{m''})(\xi_{m'} / \xi_{m''})$ 
in the regime $\xi_{m''+1} \ll L \ll \xi_{m''}$.
\autoref{fig:transmissions_2ndorder} shows the transmission probabilities $T_{m'4}$ compared
to the model predictions when these higher-order processes are also taken into account 
up to second order, showing excellent agreement.

\section{Summary}\label{sec:summary}
We have shown that in a perpendicular magnetic field 
2D nano-wires with one-sided surface 
disorder feature a regular island in phase space which leads to giant localization 
lengths in 
the limit of large Fermi momentum $\kF$, where the classical phase space structure 
can be fully resolved quantum mechanically.
The coupling between the regular island and
the chaotic sea proceeds only by tunneling, which is exponentially suppressed
in the semi-classical limit. 
Based on this understanding, we have derived analytical results for the mode-specific 
localization length $\xi_m$ in the limit of large $k_F$, which
show excellent agreement with the numerical data,
without resorting to any fit parameters.

Finally, we have presented a model describing the behavior of the transmission
probabilities $T_{mn}$ between the lowest modes $m,n$ which enter and exit the wire 
on the regular island in phase space. Taking into account how in the interior of the 
wire these modes dynamically tunnel to the counter-propagating chaotic sea and back to the
island, our model shows remarkably good agreement with the numerical data considering 
its simplicity. 

\appendix
\section{Analytical estimates of localization lengths}\label{sec:ana}
In this appendix we derive the analytical estimates of the mode localization
lengths $\xi_m$. 
The wire consists of a chain of rectangular modules, with its length given in
units of the module length $l$.

\subsection{Reduction to a single module boundary}
The transmission matrix can be constructed by connecting the transmission
matrices of the subsystems.

\begin{figure}
  \centering
  \includegraphics[width=\linewidth]{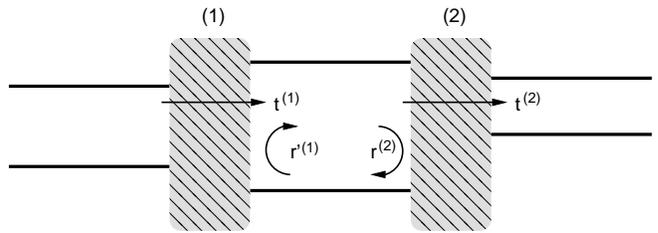}
  \caption{Connection of two scattering systems. Transmission proceeds by transmission through the
      first system $(t^{(1)})$, an arbitrary number of reflections between the two
      systems (each of which gives a term $r'^{(1)} r^{(2)}$), and then transmission
      through the second system $(t^{(2)})$, leading to \refeq{eq:tconn}.}
  \label{fig:scattconn}
\end{figure}

As a building block we first consider the connection of only two substructures such as, \eg, 
two modules (see \autoref{fig:scattconn}).
The transmission matrix $t^c$ from left to right is given by
\begin{equation}\label{eq:tconn}
  t^c = t^{(2)}\left[ \sum_{n=0}^{\infty} \left(r'^{(1)} r^{(2)} \right)^n \right]t^{(1)}\,,
\end{equation}
where $t^{(1)}, t^{(2)}$ are the transmission matrices from left to right of the two 
subsystems, $r'^{(1)}$ is the reflection matrix from the right side 
for system $1$ and $r^{(2)}$ is the reflection matrix from the left side for system $2$. 

We are interested in the transmission $t^c_{mm}$ of mode $m$ into itself for the case 
that mode $m$ is well inside the regular island. We can therefore
neglect all terms involving reflection matrices, as these would involve 
tunneling and are exponentially suppressed. This leaves 
\begin{equation}\label{eq:tcapprox}
  t^c_{mm} \approx (t^{(2)} t^{(1)})_{mm} = \sum_{n=1}^{N} t_{mn}^{(2)} t_{nm}^{(1)} 
    \approx t^{(1)}_{mm} t^{(2)}_{mm}
    \,,
\end{equation}
where $N$ is the number of modes in the module between the two systems. 
In \autoref{eq:tcapprox} we have neglected all but 
the $m$th term in that sum, as these terms involve 
tunneling to another mode and then tunneling back. 

Extending this analysis to a wire with $L$ modules we get
\begin{equation}\label{eq:tmmprod1}
  t_{mm} \approx \prod_{i=0}^L t_{mm}^{(i,i+1)}.
\end{equation}
Here, $t_{mm}^{(i,i+1)}$ is the transmission of the $m$th mode into 
itself from module $i$ to module $i+1$, with $i=0$ and $i=L+1$ labeling the 
left and right lead, respectively. From the Onsager-Casimir symmetry 
relations\cite{Ons1931,Cas1945,Bue1986} follows that $t_{mm}$ only depends 
on the two heights $h_{i}$ and $h_{i+1}$, but not on the order in which they 
occur. Additionally, for modes $m$ well inside the regular island, 
only the exponentially
suppressed tunneling tail reaches the upper side, so that the wave function in
the module with larger height $h$ can be assumed to be that of an infinitely high 
module as if  unperturbed by the upper wall. This means that only the smaller of 
the two heights will influence the transmission. We express
the small deviation of $ t_{mm}^{(i,i+1)}$ from unity
by the function $\varm[h]$, which will be related 
to the localization lengths,
\begin{equation}
  t_{mm}^{(i,i+1)} = 1 - \varm[\min(h_i,h_{i+1})]\,.
\end{equation}
Ordering the modules with increasing heights $h^{(\mu)}$, 
$\mu=1, ..., M$, we can rewrite \refeq{eq:tmmprod1}
as a product over the $M$ modules,
\begin{equation}\label{eq:tmmprod3}
 t_{mm} \approx \prod_{\mu=1}^M \left(1-\varm[h^{(\mu)}]\right)^{L P_\mu} \approx \exp(-L\sum_{\mu=1}^{M} P_\mu\varm[h^{(\mu)}])\,,
\end{equation}
where $P_\mu$ is the probability that at a module boundary
the minimal height of the adjacent modules is $h^{(\mu)}$.

We use again that the $m$th mode is exponentially suppressed at the upper
boundary, from which follows that $\varm[\hmin]$, belonging to the module
with the lowest height $\hmin\equiv h^{(1)}$, 
is the dominant contribution in \refeq{eq:tmmprod3}. 
Neglecting all other contributions and using that the probability that 
one of the modules at a module boundary is the lowest one
is $P_1=2/M$, $t_{mm}$ is simply given by
\begin{equation}\label{eq:tmmapproxeq}
 t_{mm} \approx 
 \exp\left(-\frac{2L}{M} \varm[\hmin]\right).
\end{equation}
The transmission probability $T_m\approx |t_{mm}|^2$ of mode $m$ decays according to $T_m \propto \exp(-L/\xi_m)$, 
allowing us to extract the mode localization length $\xi_m$ as
\begin{equation}\label{eq:loclenofeps}
 \xi_m \approx \frac{M}{4\varm[\hmin]}.
\end{equation}

\subsection{Transmission at the boundary of two leads}
\begin{figure}
  \centering
  \includegraphics[width=60mm]{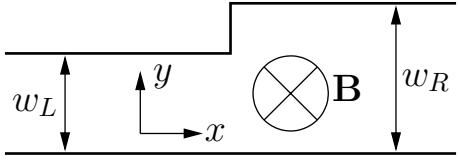}
  \caption{Two leads with widths $w_L$ and $w_R$.}
  \label{fig:twoleads}
\end{figure}

We now calculate the reduction $\varm[\hmin]$ of the transmission amplitude from 1
at a module boundary with the lower height $\hmin$.
This can be done by calculating the transmission amplitude $t_{mm}$ for the simple system of two connected leads 
with different widths $w_L$ and $w_R$ (see \autoref{fig:twoleads}).

\begin{figure}
  \centering
  \includegraphics[width=\linewidth]{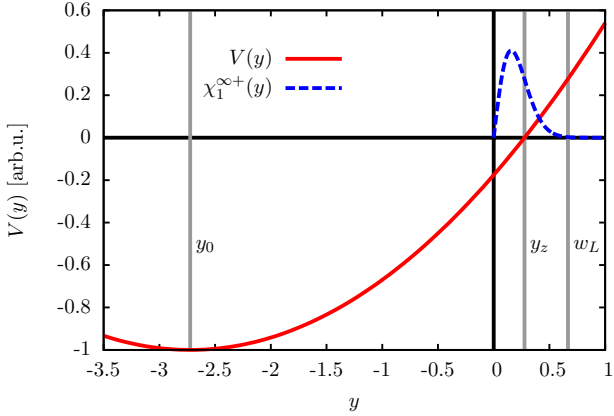}
  \caption{(Color online) Effective potential $V(y)$ and lowest mode in infinitely wide lead, for $\heff^{-1}=9.6$ and $r_c=3$.
  In addition, $w_L=\hmin$ for the geometry chosen in this paper is shown.}
  \label{fig:eff_pot}
\end{figure}

We solve the \Schro equation in each lead and 
find the transmission matrix element $t_{mm}$ by wave 
function matching at the boundary between the leads.
Inserting the Hamiltonian \autoref{eq:land_ham} into the \Schro equation and
separating the wave function as $\exp(i k_x x)\chi(y)$, we get
\begin{equation}\label{eq:effschroed}
	\left\{ \frac{p_y^2}{2} + \EF \left[\left(\frac{y-\yzero}{r_c}\right)^2 - 1 \right] \right\} \chi(y) = 0\,,
\end{equation}
which is an effective 1D-\Schro equation for a particle at energy $E=0$ in the diamagnetic potential
$V(y)=\EF[(y-\yzero)^2/r_c^2 - 1]$, where $\yzero = -k_x/B = -r_c k_x/\kF$ (see \autoref{fig:eff_pot}).

The transverse mode wave functions $\chi_m(y)$, which are zero at the boundaries $y=0$ and $y=w$, satisfy
the generalized orthogonality condition\cite{BarSto1989,SchWylRav1990}
\begin{equation}\label{eq:genorthrel}
  \Int_0^w \left(k_{x,n}^{\pm} + k_{x,m}^{\pm} + 2\kF \frac{y}{r_c}\right) \chi_n^{\pm}(y) \chi_m^{\pm}(y) \ud y = \pm\delta_{nm}\,,
\end{equation}
where the modes $\chi_m(y)$ have been normalized to carry unit flux, so that the scattering matrix is unitary.

We now assume that we have an incoming (right-moving) wave $\phi_m^{L+}$ in the $m$th mode in the left lead.
The wave functions $\Psi^L$ in the left lead and $\Psi^R$ in the right lead are given by
\begin{align}
  \Psi^L(x,y) &= \phi_m^{L+}(x,y) + \sum_{n=1}^{\infty} r_{nm} \phi_n^{L-}(x,y),\\
  \Psi^R(x,y) &= \sum_{n=1}^{\infty} t_{nm} \phi_n^{R+}(x,y),\\
  \intertext{where}
  \phi_n^{S\pm}(x,y) &= \chinorm{n}{S\pm} \exp\left(i k_{x,n}^{S\pm} x\right).
\end{align}

The continuity condition for the wave functions at $x=0$ demands that 
\begin{align}\label{eq:wfmatch}
  \Psi^L(x,y)\atxz &= \Psi^R(x,y)\atxz,\\
\intertext{and}
\label{eq:wfmatchderiv}
  \frac{\partial\Psi^L}{\partial x}(x,y)\atxz &= \frac{\partial\Psi^R}{\partial x}(x,y)\atxz,
\end{align}

In order to extract $t_{mm}$, we multiply \refeq{eq:wfmatch} by
$(\kxsh[R+]+2\kF y/r_c)$, add to it $(-i)$ times \refeq{eq:wfmatchderiv},
multiply this equation by $\chinorm{m}{R+}$, and integrate from $y=0$
to $y=\infty$. The right hand side reduces to the generalized
orthogonality relation (\ref{eq:genorthrel}) for the transverse wave functions
in the lead and therefore simplifies to $t_{mm}$,
\begin{equation}\label{eq:tmmtwoleads}
    t_{mm} = O^{m+}_{m+} +  \sum_{n=1}^{\infty} r_{nm} O^{n-}_{m+} \,,
\end{equation}
with $O^{n_L\pm}_{n_R\pm}$ being the generalized overlap integral between mode $n_L$ in the left lead and mode
$n_R$ in the right lead,
\begin{equation}
O^{n_L\pm}_{n_R\pm} = \Int_0^{w_<} \left(k_{x,n_L}^{L\pm}+k_{x,n_R}^{R\pm}+2\kF \frac{y}{r_c}\right) \chinorm{n_L}{L\pm}\chinorm{n_R}{R\pm} \ud y \,,
\end{equation}
where $w_<$ is the smaller of the two lead widths $w_L$, $w_R$. For
sufficiently high magnetic field $B$ and Fermi momentum $\kF$, the wave
functions in the $m$th mode are small at the upper boundary, so that we expect
$t_{mm}$ to be almost one. We define its deviation from 1 by
\begin{equation}\label{eq:varm}
        t_{mm} = 1 - \varm \,,
\end{equation} 
where $\varm$ for two leads corresponds to $\varm[h_1]$ of the rough wire introduced in the previous subsection.

Inserting \refeq{eq:varm} into the unitarity condition
\begin{align}
  1 &= \sum_{n=1}^{N} \left( \abs{t_{nm}}^2 + \abs{r_{nm}}^2 \right),\\
  2\varm &\approx \sum_{n\ne m} \abs{t_{nm}}^2 + \sum_{n=1}^{N} \abs{r_{nm}}^2 \,,
\end{align}
we find that the $r_{nm}$ can be at most $O(\sqrt{\varm})$.
The second term in \refeq{eq:tmmtwoleads} approximately corresponds to the orthogonality condition
(\ref{eq:genorthrel}) for $n \not= m$, so that it is strongly suppressed. The integrals can be estimated to
be of order $O(\sqrt{\varm})$ by using that the difference between the $m$th
modes on the left and right side is of order $O(\sqrt{\varm})$, while the
left-moving modes are of order $O(1)$ at the upper side of the wire, where
$\chinorm{m}{R+}$ differs from $\chinorm{m}{L+}$. Utilizing that the integral and
$r_{nm}$ are both $O(\sqrt{\varm})$, the whole term should be of order
$O(\varm)$, such that a priori it cannot be neglected. 
Numerically, we find that its magnitude 
does not exceed $0.2 \varm$, quite independently of the Fermi energy
$\EF$ and the magnetic field $B$. We neglect the second term in \refeq{eq:tmmtwoleads} in the following,
keeping in mind that this will introduce an error of about $20\%$ in our
result for the localization lengths. 

\begin{figure}
  \centering
  \includegraphics[width=\linewidth]{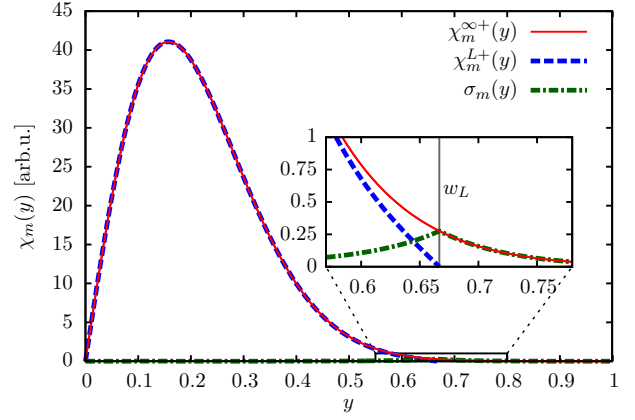}
	\caption{(Color online) Sketch of the first mode ($m=1$) transverse wave function $\chinorm{m}{L+}$ in a lead of width $w_L$,
	  $\chish$ in an infinite lead, and the difference $\sigma_m(y)$, for $1/\heff=9.6$ and $r_c=3$.
	  The inset shows an enlargement of the region around $y=w_L$.}
	\label{fig:wavfunsigma}
\end{figure}

We thus approximate
\begin{equation}\label{eq:tmmtwoleadsapprox}
    t_{mm} 
    \approx
    \Int_0^{w_<} \left(k_{x,m}^{L+}+k_{x,m}^{R+}+2\kF \frac{y}{r_c}\right) 
    \chinorm{m}{L+}\chinorm{m}{R+} \ud y\,,
\end{equation}
which is independent of the order of the leads. Without loss of generality,
we can choose $w_L < w_R$ for the further calculation.
Since the $m$th mode in the (wider) right lead is much less affected by the 
upper boundary than the $m$th mode in the (narrower) left lead, we replace the former 
by the wave function of the $m$th mode in an infinitely wide lead,
\begin{equation}\label{eq:chiright}
  \chinorm{m}{R+} \approx \chish.
\end{equation}
The transverse wave function of the $m$th mode in the left lead differs from this only
very slightly, see \autoref{fig:wavfunsigma}, so we write it as
\begin{equation}\label{eq:chileft}
  \chinorm{m}{L+} = N' \left[\chish - \sigmy\right],
\end{equation}
where $\sigmy$ is negligible except near $y\!=\!w_L$ and is given by
$\sigmy = \chish$ for $y>w_L$, where $\chinorm{m}{L+}=0$.
The normalization factor $N'$ is close to one and will be evaluated below. 
The longitudinal momentum in the left lead can be written as
\begin{equation}\label{eq:kxdelta}
  \kxsh[L+] = \kxsh + \Dkxm.
\end{equation}
The difference $\Dkxm$ is very small and is
numerically found to be $O(\varm)$. 

Inserting (\ref{eq:chiright}-\ref{eq:kxdelta}) into \refeq{eq:tmmtwoleadsapprox} and extending the integral to
$y\!=\!\infty$ (considering that the wave function $\chinorm{m}{L+}$
is zero for $y>w_L$), we have 
\begin{multline}
  t_{mm}= \Int_0^{\infty} \left(2 \kxsh +\Dkxm +2\kF \frac{y}{r_c}\right)\\
  \qquad N' \left[\chish - \sigmy\right] \chish \,\ud y \,,
\end{multline}
and thus
\begin{equation}\label{eq:tmm_intapprox}
  t_{mm} = N' \left[ 1 - A + \Dkxm B \right] \,,
\end{equation}
with
\begin{align}
  A  &= \Int_0^{\infty} \left(2 \kxsh + 2\kF \frac{y}{r_c}\right) \sigmy\chish \ud y,\\
  B  &= \Int_0^{\infty} \left[\chish-\sigmy\right]\chish \ud y,
\end{align}
To calculate $N'$, we use the flux normalization condition from \refeq{eq:genorthrel},
\begin{multline}
  1 = N'^2 \Int_0^{\infty} \left(2 \kxsh + 2\Dkxm + 2\kF \frac{y}{r_c}\right) \\
  \qquad \left[\chish-\sigmy\right]^2\,\ud y,
\end{multline}
which we write as
\begin{equation}
  1  = N'^2 \left[ 1 - 2A + C + 2\Dkxm B' \right],
\end{equation}
with
\begin{align}
  B' &= \Int_0^{\infty} \left[\chish-\sigmy\right]^2 \ud y,\\
  C  &= \Int_0^{\infty} \left(2 \kxsh + 2\kF \frac{y}{r_c}\right) \sigmy^2 \ud y.
\end{align}
Inserting $N'$ into \autoref{eq:tmm_intapprox}, we obtain
\begin{align}
  t_{mm} &= \frac{1 - A + \Dkxm B}{\sqrt{1 - 2A + 2\Dkxm B' + C}} \\
  &\approx 1 - \frac{1}{2} C + \Dkxm \left(B - B'\right)\,,
\end{align}
where we used that the integrals $A,B,B',C$ are much smaller than one.
With the explicit expressions for $B$ and $B'$, the last term becomes
\begin{equation}
 \Dkxm (B-B') = \Dkxm \Int_0^{\infty} \left[\chish\sigmy - \sigmy^2\right] \ud y,
\end{equation}
which is of higher order than integral $C$ since $\sigmy$ is almost zero
where the wave function has its maximum and $\Dkxm$ is already
$O(\varm)$. Dropping this term, we arrive at a simple
expression for $\varm$,
\begin{equation}\label{eq:epsapproxfin}
    \varm \approx \frac{1}{2}C = \Int_0^{\infty} \left(\kxsh + \kF \frac{y}{r_c}\right) \sigmy^2 \ud y.
\end{equation}

\subsubsection{WKB approximation of \texorpdfstring{$\sigma_m(y)$}{sigma\_m(y)}}

To evaluate \autoref{eq:epsapproxfin}, we need an expression for the function $\sigmy$, see \autoref{fig:wavfunsigma}.  By
inserting \refeq{eq:chileft} into \refeq{eq:effschroed} and imposing the
boundary conditions $\chinorm[0]{m}{L+} = \chinorm[w_L]{m}{L+} = 0$, we
find that $\sigmy$ has to be an eigenfunction of the same Hamiltonian
$H^{y}$ as $\chinorm{m}{L+}$, but with boundary conditions $\sigmy[0] =
\sigmy[\infty] = 0$. Instead of a normalization condition, it has to satisfy 
$\sigmy[w_L] = \chish[w_L]$. Since the
upper boundary $w_L$ is already deep in the classically forbidden region, we
use a WKB approximation for our solution for $\sigmy$,
\begin{multline}
  \sigmy = \chish[w_L] \sqrt{\frac{\rho_m(w_L)}{\rho_m(y)}} \times \\
  \times \begin{cases}
            \exp\left(-\Int_{y}^{w_L} \rho_m(y') \ud y'\right) & y<w_L\\[1em]
            \exp\left(-\Int_{w_L}^{y} \rho_m(y') \ud y'\right) & y>w_L
          \end{cases},
\end{multline}
with
\begin{equation}
  \label{eq:rhodef}
  \rho_m(y) = \sqrt{2[V_m(y)-E]} = \kF \sqrt{V_m(y)/\EF},
\end{equation}
where we have used that the effective 1D \Schro equation has eigenvalue
$E=0$ and the effective potential $V_m$ depends on the longitudinal wave 
number $k_{x,m}$ (\cf \refeq{eq:effschroed}). Inserting the above 
expression into \refeq{eq:epsapproxfin}, we find
\begin{multline}
  \varm =
  \left[\chimwl\right]^2 \times\\
  \times \left[ \Int_{0}^{w_L} \left(\kxsh + \kF \frac{y}{r_c}\right) 
    \frac{\exp\left(-2\Int_{y}^{w_L} \rho_m(y') \ud y'\right)}{\rho_m(y)/\rho_m(w_L)} \ud y \,+ \right. \\
  + \left. \Int_{w_L}^{\infty} \left(\kxsh + \kF \frac{y}{r_c}\right)
    \frac{\exp\left(-2\Int_{w_L}^{y} \rho_m(y') \ud y'\right)}{\rho_m(y)/\rho_m(w_L)} \ud y \right].
\end{multline}
We extend the first integral to start at $-\infty$ and perform 
in the first  integral the
substitutions $\eta'=w_L-y'$, $\eta=w_L-y$ and $\eta'=y'-w_L$, $\eta=y-w_L$
in the second.
Since the integrals will only give a significant contribution near $y=w_L$,
\ie $\eta=0$, we expand $\rho_m(y)$ into a Taylor series to first order,
\begin{equation}
        \rho_m(w_L\!+\!\eta) \approx \rho_m(w_L) + \rho'_m(w_L) \eta.
\end{equation}
Defining $\rho_m\equiv\rho_m(w_L)$, $\rho'_m\equiv\rho'_m(w_L)$ and
$\chi_m^{\infty+}\equiv\chimwl$, we find
\begin{multline}
  \varm = \left(\chimnoarg\right)^2 \Int_{0}^{\infty} \ud \eta \exp\left(-2\rho_m \eta\right) \times \\
  \times \left[ \left(\kxsh + \kF\frac{w_L - \eta}{r_c} \right) \frac{\exp\left(\rho'_m \eta^2\right)}{1 - \eta\rho'_m/\rho_m} \right. + \\
  + \left. \left(\kxsh + \kF\frac{w_L + \eta}{r_c} \right) \frac{\exp\left(-\rho'_m \eta^2\right)}{1 + \eta\rho'_m/\rho_m} \right].
\end{multline}
Expanding the term in square brackets in powers of $\eta$ gives
\begin{equation}
   [\ldots] = 2\left(\kxsh + \frac{\kF w_L}{r_c}\right) + O(\eta^2).
\end{equation}
Dropping the quadratic term, the evaluation of the integral leads to 
\begin{equation}\label{eq:epsawkb}
        \varm = \frac{\left(\chimnoarg\right)^2}{\rho_m} 
                \left(\kxsh + \frac{\kF w_L}{r_c}\right),
\end{equation}
with $\chimnoarg$ and $\rho_m$ both evaluated at $y\!=\!w_L$.

\subsubsection{WKB approximation of \texorpdfstring{modes $\chish$}{lead modes}}

The next step
in our calculation of $\varm$, and, ultimately, of the mode localization
length $\xi_m$, is to find an expression for the value $\chimwl$ of the
transverse wave function at the upper boundary and for the
longitudinal momentum eigenvalue $\kxsh$.

To do this, we rewrite the effective transverse Hamiltonian $H^y$, following
from \refeq{eq:effschroed},
\begin{equation}
	H^y = \frac{p_y^2}{2} + V(y),
	\label{eq:efftranshamil}
\end{equation}
with the potential
\begin{equation} 
  V(y) = \frac{\kF^2}{2} \left[ \left(\frac{y-y_z}{r_c}\right)^2 + 2\frac{y-y_z}{r_c} \right],
\end{equation}
with the classical turning point $y_z = \yzero + r_c$,
where the potential value $V(y_z) = 0$ equals the energy $E=0$. We linearize the potential near the classical
turning point (where the WKB solution diverges), which makes it possible 
to solve the effective \Schro equation analytically. The complete wave function is
constructed by connecting the solution of the linearized potential near
the classical turning point to the WKB solution in the classically forbidden
region.

The linearized \Schro equation is solved by the Airy function $\Ai(z)$
with $z=(2\kF^{2}/r_c)^{1/3}(y-y_z)$. The boundary condition at the lower wall demands 
that $\chi(y\!=\!0)$ is equal to zero, so that $z(y\!=\!0)$ must be a zero of $\Ai(z)$.
For the solution for the $m$th mode, we choose the $m$th zero at
$z=\aizm\approx(-2.338,-4.088,-5.521,-6.787,\dots)$, so that the wave function has $m-1$ nodes.
From this follows that
\begin{align}
  y_z &= -\left(\frac{r_c}{2\kF^{2}}\right)^{1/3} \aizm, \label{eq:yzsol}\\
  \intertext{and by using $\yzero = -r_c k_x/\kF$ we find}
  \kxsh &= \kF \left[ 1 + \frac{\aizm}{2^{1/3}} \left(\kF r_c\right)^{-2/3} \right].
\end{align}
For large $\kF$ the longitudinal momentum of the $m$th mode $\kxsh$ is only
marginally smaller than the Fermi momentum $\kF$ (remember
that $\aizm$ is negative). The maximum of the $m$th transverse mode is between
$z=\aizm$ and $z=0$, \ie between $y=0$ and $y=y_z$.
Since $y_z$ approaches $y=0$ for large $\kF$, the wave function stays closer
and closer to the lower wall with increasing $\kF$.

The transverse wave function near the classical turning point can now be written as
\begin{equation}\label{eq:chiairysol}
  \chish = C_m \Ai\left(\aizm + \left(\frac{2\kF^{2}}{r_c}\right)^{1/3} y\right).
\end{equation}
We will construct the full solution for the wave function by using the Airy
function near the classical turning point and the WKB solution (which takes
the quadratic potential into account) in the classically forbidden region.
Before proceeding, we determine the prefactor $C_m$. Since the WKB solution is
only used in describing the exponential tail for $y\gg y_z$, calculating $C_m$
with the wave function of the linearized potential will
only introduce a small error. Therefore, we insert \refeq{eq:chiairysol} into
the flux normalization condition \refeq{eq:genorthrel} and obtain
\begin{equation}
  1 = \Int_0^\infty \left(2\kxsh + 2\kF \frac{y}{r_c}\right) \left[\chinorm{m}{\infty+}\right]^2 \ud y,
\end{equation}
which results in
\begin{equation}
  C_m = \left[(4\kF r_c)^{1/3} \left(\aiintam + \frac{\aiintbm}{2^{1/3}\left(\kF r_c\right)^{2/3}} \right)\right]^{-1/2},
\end{equation}
where $\aiintam = \Int_{\aizm}^\infty \Ai^2(z) \ud z$ and $\aiintbm = \Int_{\aizm}^\infty z\Ai^2(z)\ud z$.
In the limit of large $\kF$, this simplifies to
\begin{equation}\label{eq:c1appr}
  C_m \approx \left[(4\kF r_c)^{1/3} \aiintam \right]^{-1/2}.
\end{equation}

Since we need to evaluate the transverse wave function $\chimnoarg$ at
$y\!=\!w_L$, which is deep in the classically forbidden region, we proceed by
connecting the Airy function (valid near the classical turning point) to the
WKB solution (valid in the classically forbidden region). We write the WKB
solution as
\begin{equation}\label{eq:chiasymwkb1}
  \chish \approx \frac{D_m}{\sqrt{\rho_m(y)}} \exp\left(-\Int_{y_z}^{y} \rho_m(y') \ud y' \right) \qquad y \gg y_z,
\end{equation}
and from a short calculation we obtain that the two constants $C_m$ and $D_m$ are related by
\begin{equation}\label{eq:c2c1conn}
  D_m = \left(\frac{2\kF^{2}}{r_c}\right)^{1/6}\frac{C_m}{2\sqrt{\pi}}.
\end{equation}

For evaluating the integral in \refeq{eq:chiasymwkb1}, we insert the explicit
form of the potential, use $y\!=\!w_L$ as the upper limit of
integration, and rewrite \refeq{eq:rhodef} as
\begin{equation}
    \rho_m(y) = \kF \sqrt{2\frac{y-y_z}{r_c} + \left(\frac{y-y_z}{r_c}\right)^2},
\end{equation}
leading to
\begin{equation}
  \Int_{y_z}^{w_L} \rho_m(y') \ud y' = \kF r_c \Int_0^{\bar z} \sqrt{2z' + z'^2} \ud z',
\end{equation}
with ${\bar z}=(w_L-y_z)/r_c$\,. We evaluate the above integral by expanding the integrand in powers of $z'$,
\begin{align}
  \Int_{y_z}^{w_L} \rho_m(y') \ud y' &= \sqrt{2} \kF r_c \Int_0^{\bar z} \left(z'^{1/2} + \frac{z'^{3/2}}{4} + O(z^{5/2})\right) \ud z'\nonumber\\
   &\approx \frac{2\sqrt{2}}{3} \kF r_c {\bar z}^{3/2} \left(1 +  \frac{3}{20} {\bar z}\right).
\end{align}

Inserting this into \refeq{eq:chiasymwkb1}, and using that $y_z \ll w_L$ in the limit of large $\kF$, we obtain
\begin{multline}\label{eq:chiasymwkb2}
  \chish[w_L] \approx \frac{D_m}{\sqrt{\rho_m(w_L)}} \exp\left[-\frac{2\sqrt{2}}{3} \kF r_c \right. \\
  		                \left. \left(\frac{w_L}{r_c}\right)^{3/2} \left(1-\frac{y_z}{w_L}\right)^{3/2} \left(1 + \frac{3}{20} \frac{w_L}{r_c}\right) \right],
\end{multline}
neglecting higher order terms.

We now insert this into \refeq{eq:epsawkb} and obtain
  \begin{multline}\label{eq:epsawkbfinal}
    \varm = E_m \exp\left[-\frac{4\sqrt{2}}{3} \kF r_c \left(\frac{w_L}{r_c}\right)^{3/2} \right.\\
    	         \left. \left(1 + \frac{3}{20} \frac{w_L}{r_c}\right) \left(1-\frac{y_z}{w_L}\right)^{3/2} \right],
  \end{multline}
with the prefactor
  \begin{equation}
    E_m =\frac{(\kF r_c)^{-2/3}}{2^{1/3} 4\pi \aiintam} \,
        \frac{\kxsh/\kF + w_L/r_c}{(w_L-\yzero)^2/r_c^2 -1} \,,
  \end{equation}
where we used \refeq{eq:c2c1conn}, \refeq{eq:c1appr}, and \refeq{eq:rhodef}.

\subsection{Semiclassical limit with constant cyclotron radius}
To obtain our final result for the mode localization length $\xi_m$ in the
semiclassical limit of large $\kF$, we insert 
\refeq{eq:epsawkbfinal} with $w_L=\hmin$ 
into the expression \refeq{eq:loclenofeps} for $\xi_m$.
We keep the cyclotron radius $r_c = c \kF/B$ constant, 
such that the classical dynamics is independent of $\kF$. 
We expand the prefactor $M/(4 E_m)$ in powers of $\kF$ for $\kF\to\infty$, keeping the first 
two terms in the expansion since they are of similar magnitude for the parameter values used,
and finally obtain
\begin{equation}\label{eq:osd_anarcconstloclenpred2}
   \xi = \left(a_m \heff^{-2/3} - b_m\right) \exp\left[ c_0 \heff^{-1} (1-d_m \heff^{2/3})^{3/2}\right],
\end{equation}
where $\heff = (\kF W/\pi)^{-1}$ and the dimensionless parameters $a_m$, $b_m$, $c_0$, and $d_m$ are given by
\newpage
\begin{subequations}
  \begin{align}
    a_m &= 2^{1/3} \pi^{5/3} M \aiintam \frac{\Delta}{\zeta^{1/3}} \left(\frac{2 + \Delta/\zeta}{1+\Delta/\zeta}\right),\\
    b_m &= -\pi \aizm M \aiintam \left( 1 + \frac{\zeta^2}{(\Delta+\zeta)^2} \right),\\
    c_0 &= \frac{4\sqrt{2}\pi}{3} \frac{\Delta^{3/2}}{\zeta^{1/2}} \left(1+\frac{3}{20}\frac{\Delta}{\zeta}\right),\\
    d_m &= \frac{-\aizm}{2^{1/3} \pi^{2/3}} \frac{\zeta^{1/3}}{\Delta},
  \end{align}
\end{subequations}
where we have introduced the dimensionless parameters $\Delta = \hmin/W = 1-(\delta/2W)$ 
and $\zeta=r_c/W$. 
Note that the factor $c_0$ determining the asymptotic exponential behavior does not depend on
the mode number $m$.
For $m=1$ and in the limit $\Delta/\zeta \ll 1$ this corresponds to the result
we have previously reported.\cite{FeiBaeKet2006}

\subsection{Semiclassical limit with constant magnetic field}
Instead of keeping the cyclotron radius fixed, we alternatively set the magnetic field $B$ to a
fixed value independent of $\kF$ and again perform the limit $\kF\to\infty$. 
We repeat the above procedure, introducing the magnetic length $\lambda$ through $B/c = 1/\lambda^2$
and the dimensionless parameter $\Lambda = \lambda/W$. As above, we use $\Delta = \hmin/W$ and obtain
\begin{equation}\label{eq:osd_anabconstloclenpred2}
  \xi_m = \left( a_m \heff^{-1/3} - b_m \right) \exp( c_0 \heff^{-1/2}
  (1-d_m \heff^{1/3})^{3/2}),
\end{equation}
with the dimensionless parameters $a_m$, $b_m$, $c_0$, $d_m$ now given by
\begin{subequations}
  \begin{align}
    a_m &= (2\pi)^{4/3} M \aiintam \frac{\Delta}{\Lambda^{2/3}},\\
    b_m &= -2\pi \aizm M \aiintam,\\
    c_0 &= \frac{4\sqrt{2\pi}}{3} \frac{\Delta^{3/2}}{\Lambda},\\
    d_m &= \frac{-\aizm}{(2\pi)^{1/3}}\frac{\Lambda^{2/3}}{\Delta} .
  \end{align}
\end{subequations}

\end{document}